\documentclass[prd,twocolumn,amsmath,nobibnotes,superscriptaddress]{revtex4}
\usepackage{bm} \usepackage{graphicx} \usepackage{epstopdf}
 \def\ba{\begin{eqnarray}} \def\ea{\end{eqnarray}}
\def\be{\begin{equation}} \def\ee{\end{equation}} \def\({\left(}
\def\){\right)} \def\[{\left[} \def\]{\right]} \def\<{\left<}
\def\>{\right>}

\newcommand{\rar}{\rightarrow}

\newcommand{\mpl}{\ensuremath{M_\text{Pl}}}

\begin{document}

\title{Hurdles for Recent Measures in Eternal Inflation}
\date{\today}

\author{Anthony Aguirre}
\email{aguirre@scipp.ucsc.edu}
\affiliation{SCIPP, University of California, Santa Cruz, CA 95064, USA}
\author{Steven Gratton}
\email{stg20@cam.ac.uk}
\affiliation{Institute of Astronomy, Madingley
 Road, Cambridge, CB3 0HA, UK}
\author{Matthew C Johnson}
\email{mjohnson@physics.ucsc.edu}
\affiliation{SCIPP, University of California, Santa Cruz, CA 95064, USA}

\begin{abstract}
In recent literature on eternal inflation, a number of measures have
been introduced which attempt to assign probabilities to different
pocket universes by counting the number of each type of pocket
according to a specific procedure. We give an overview of the existing
measures, pointing out some interesting connections and generic
predictions. For example, {\em pairs} of vacua that undergo fast
transitions between themselves will be strongly favored. The resultant
implications for  
making predictions in a generic potential landscape are discussed. We
also raise a number of issues concerning the types of transitions that
observers in eternal inflation are able to experience.     
\end{abstract}

\maketitle

\section{Introduction}

In eternal inflation, different post-inflationary regions may have
different properties.  How -- even in principle -- to statistically
describe these properties so as to make probabilistic cosmological
predictions is a major outstanding problem in current cosmology.
Recently, a number of proposals have been advanced for
``gauge-independent" measures that do not depend on the choice of a
time  coordinate~\cite{Bousso:2006ev,Easther:2005wi,Garriga:2005av,Vanchurin:2006qp}.  In this note we compare, contrast, and assess the existing proposals, and point out some predictions that they seem to share.  We focus here on eternal inflation as driven
by a potential with multiple minima; transitions between these correspond to the nucleation of bubbles or ``pocket universes" containing a new phase of different vacuum energy~\cite{Coleman:1980aw,Lee:1987qc}. If transitions are sufficiently slow, the growing bubbles never percolate, and inflation is eternal.

A form of predictions in a multiverse is a set of statements such as
``The probability that a randomly chosen $X$ is in a region with
properties $\alpha$ is ${\cal P}_X(\alpha)$", where $X$ is some
``conditionalization object" such as a point in space, a baryon, a
galaxy, or an ``observer" that arguably makes ${\cal P}_X$ relevant to
what we will actually observe in some future experiment (see,
e.g.,~\cite{Aguirre:2004qb,Aguirre:2005cj}).  This probability is
generally split into two components:
\begin{equation}
\label{eq-psubx}
{\cal P}_{X}(\alpha) \propto P_p(\alpha) n_{X,p}(\alpha).
\end{equation}
Here, $P_p$ is a ``prior" probability distribution defined in terms of
some type of object $p$ regardless of the conditionalization object
$X$, and $\alpha$ is a vector of properties we might hope to compare to
locally observed properties of our universe. For example, if
$p$=``pocket universe" then $P_p(\alpha)$ describes the probability
that a randomly chosen bubble has low-energy observable properties
$\alpha$. The factor $n_{X,p}$ conditions these probabilities by the
requirement that some $X$-object exists; for example with
$X$=``galaxy", $n_{X,p}(\alpha)$ might count the ($\alpha$-dependent)
number of galaxies in a pocket with properties $\alpha$.

The measures discussed here are proposals for calculating $P_p$
(though we will also discuss some relevant issues concerning
$n_{X,p}$).  We shall see that many of the measures share some
properties -- for example, they all accord very high probability to
regions in a potential landscape which allow for very rapid
transitions between nearby minima. Unfortunately, these regions of the
landscape look nothing like our universe: the resulting spacetimes
would almost certainly be dominated by a very high vacuum energy and
be devoid of structure.  This, of course is nothing new -- the whole idea
of the ``anthropic" approach to explaining our observed universe is
that $n_{X,p}$, where $X=$``observer", will ``unweight" such states.
But we shall see that employing the measures under consideration makes
the problem very acute.
 
   More generally, while the measures we discuss are all stated
and formulated in rather different ways, many of them are, in fact,
either fully or partially equivalent (as acknowledged by the authors
in some cases); we will attempt to sort out these relations
comprehensively.  Differences do exist however, and we will also see
that certain ``desirable" properties hold in some measures and not
others.

In Section~\ref{Counting}, we present a ``scorecard'' of features that
might be desirable in a measure, and we summarize a number of recent
measures and the connections between them. We then compute the prior
distribution for a number of sample landscapes in Sec.~\ref{samples} and
use the results to highlight important predictions and
connections. The implications of these predictions are discussed in
Sec.~\ref{consequences}. Sec.~\ref{observers} describes some problems
associated with the assumptions usually made about the global picture
of an eternally inflating spacetime, and we conclude in
Sec.~\ref{conclusions}. 
In appendix~\ref{snowmensection}, a quick matrix method for
calculating bubble abundances is introduced and a number of the more
technical results of the paper are derived. 

\section{Measure desiderata and proposals}\label{Counting}

\subsection{Desirable measure properties: a scorecard}\label{wishlist}

To test a theory of eternal inflation yielding diverse
post-inflationary predictions, we would like to know ``what
physical properties are most likely", and compare them to our local
observations. This question, however, is simply ambiguous -- any
answerable version of this question will entail a tacit choice of a
conditionalization $X$ and calculation of ${\cal P}_X$ as described
above. The measures we will discuss correspond to different
attempts to (at least implicitly) propose a plausible candidate for
$X$, and to calculate the prior distribution $P_p$ that might be used in calculating ${\cal P}_X$ for that $X$.

A fundamental property that a well-defined measure should have is that
its answer should be gauge-invariant, by which we simply mean that its
answer can be calculated in any coordinate system we choose. This is distinct from ``gauge-independence'' as we shall discuss shortly.

Beyond this, it is important to consider what properties we might want
a sensible measure to have.  Some such desiderata, either stressed
previously in the literature or first mentioned here, are given below.
We note, however, that it is quite possible that the ``correct"
measure (if it exists) does not satisfy every item.

\begin{itemize} 

\item Physicality -- The $p$ to which the measure applies, and the choice of $P_p$, should be such that (a) the probabilities do not appear to have been ``picked out of a hat,'' and (b) $n_{X,p}$ 
 is plausibly calculable.  For example, we might choose $p=$``vacuum'' and set $P_p$ proportional to the tenth power of the hyperbolic tangent of the energy of the vacuum in Planck units. However, (a) this measure is obviously rather arbitrary, and (b) since there is no physical process behind the creation of regions described by the different vacua, the measure seems useless in calculating $n_{X,p}$ for, say $X$=``baryon." Note, however, that 
different physically reasonable conditionalization objects may require
different $P_p$ -- for example were $X$=``vacuum", then the measure would still violate condition (a), but would satisfy condition (b) by definition.

\item Gauge-independence -- The relative probabilities should not
depend on an arbitrary decomposition of spacetime into space and
time. For instance, it has been shown~\cite{Linde:1993xx,
Winitzki:2005ya,Guth:2000ka,Tegmark:2004qd} that measures that weight
based on the physical volume in a given state at late times give a
result that depends sensitively on the assumed foliation of spacetime into
equal-time hypersurfaces. In the absence of a strong physical
reason for choosing a particular decomposition, such measures thus
    seem ambiguous.
  
\item Ability to cope with varieties of transitions and vacua -- The
  measure should be general enough to treat all of the types of vacua
  (e.g.\  positive, negative, or zero energy), and the various types of
  transitions between them.  

\item Independence of initial conditions -- It is often argued that
eternal inflation approaches a steady-state, and that essentially all
observers exist ``at late times," so a physically reasonable measure
should become independent of initial conditions.  This criterion is
not obviously necessary; although it may be appropriate for a
particular conditionalization object (e.g.\ $X$=``a randomly chosen
observer''), it may not be appropriate for others. For example, if one were interested in knowing what a {\em given} observer (or worldline) will experience
in the future, then a dependence on initial conditions seems quite 
reasonable. 

\item Ability to cope with various and/or varying topological
  structures -- The measure should  potentially be applicable to spacetimes with non-trivial topological structures as may arise in eternal inflation (as discussed at length
in Sec.~\ref{observers}).

\item {Accurate and robust treatment of ``states" and ``transitions" -- this entails several sub-criteria:

\begin{itemize}

\item General principles -- the basic ideas behind the measure should allow it to be used (in theory) for the complicated ``spacetimes'' of landscapes that cannot simply be
  encapsulated by transition rates between vacua.

\item Physical description of transitions -- transition rates must be clearly linked to the physical process that describes the transition (e.g.\  Coleman-De Luccia bubble nucleation).

\item Reasonable treatment of ``split" states --  the measure 
should deal properly with very similar states and/or very large transition rates. (For example, a vacuum split by the insertion of a small potential barrier should, in the limit of an infinitesimal barrier, act just as a single vacuum.)

\item Continuity in transition rates -- When transition rates are
  used, the measure should be continuous in these rates.  For example,
  there should be no discontinuity in the probabilities between a
  stable vacuum and a metastable vacuum with a lifetime $\tau$, in the
  limit $\tau\rightarrow\infty$.

\end{itemize}
}

\end{itemize}

We would argue that all of these potentially pleasing features are
absent in at least one measure proposal in the literature, and that no
extant proposal clearly fulfills them all. But the good news is that
the bubble-counting procedures discussed here satisfy many of them, so
let us summarize these measures and provide a listing of connections
between them.

\subsection{The Measures and their Properties}\label{measprops}
 
We now examine the various measures under consideration. All of these
have subtleties, so we refer the reader to the original papers, and
also to the review by Vilenkin~\cite{Vilenkin:2006xv} and to the lectures of 
Shenker~\cite{Shenker:06}. Here, we will mainly provide brief
summaries, but will also add extended comments on some measures.

Restricting the discussion to eternal inflation as driven
by a potential with multiple minima, it is useful to classify vacua as ``terminal" or ``recycling": terminal vacua can be reached, but never exited; recycling vacua can exit to the state from which they originated, and may also transition
to other states.  Following~\cite{Bousso:2006ev}, we can also label entire landscapes as terminal or recycling; the former contain at least one terminal vacuum whereas the latter do not.  

As a first step in this analysis, we can divide the measures into
three categories: first, those that calculate volumes in different
vacua on some equal-time surface; second, those that count individual
bubbles; third, those that focus on the vacua experienced by an
 observer following a single worldline.

There are two basic volume-counting methods, counting either physical
volume (i.e.\ $p$=``unit of physical volume") or comoving volume
($p$=``unit of comoving volume"). See, e.g.,
~\cite{Linde:1993xx,Winitzki:2005ya,Guth:2000ka,Tegmark:2004qd,Linde:2006nw} for the former; here we focus on:

\begin{itemize}

\item{\em The Comoving Volume (CV) method:} Put forward by Garriga and
Vilenkin~\cite{Garriga:2001ri}, this method might be considered the
counterpart for bubble nucleations (in comoving volume) to the work of
Linde, Linde and Mezhlumian~\cite{Linde:1993xx} in stochastic
inflation.  One starts with some region on an initial spacelike surface, and considers a congruence of hypersurface-orthogonal
geodesics (the ``comoving observers'') emanating from that region.  As
a function of some global time coordinate $t$, the number of
worldlines (to which the comoving volume fraction is defined to be
proportional) in different vacua is calculated. The probability,
$P^{\text{cv}}$, to be in a given vacuum is then defined to be proportional
to the fraction of comoving volume (or number of worldlines),
$f_{i}(t)$, in that pocket, in the $t\rightarrow \infty$ limit. Note
that if there are terminal vacua, then as $t \rightarrow \infty$ all
of the comoving volume will be distributed among the terminal vacua,
except for a set of measure zero (albeit one that corresponds to
infinite physical volume!).  Metastable vacua are thus accorded zero
weight. This measure depends heavily on initial conditions, because the fraction of comoving volume in a given terminal vacuum can only increase with
time~\footnote{Note that this is worse than it may sound, because the
{\em same spacetime} might be sliced with different initial surfaces
so as to lead to completely different probability distributions.}.

\end{itemize}

The next two methods, rather than counting total relative volume in
different bubble types, count relative total {\em numbers} of bubbles,
i.e.\ $p$=``bubble".

\begin{itemize}

\item {\em The Comoving Horizon Cutoff (CHC) method}: \ \ In the
proposal of Garriga et al.~\cite{Garriga:2005av}, the measure is
defined by directly counting bubbles of a given phase. One has in mind performing the count at late times, or ``future infinity''.  We follow the most recent description of this procedure as given by
Vilenkin~\cite{Vilenkin:2006xv}. First, just as in the CV method, a
spacelike hypersurface in the spacetime is chosen, and a congruence of
geodesics is extended from this hypersurface. The geodesics are
followed arbitrarily far into the future, passing into any bubbles they may encounter. These lines are used to project bubbles in the spacetime back onto the initial hypersurface as ``colored shadows''. The relative frequency of bubbles of different colors is defined to be the ratios of the numbers of their shadows on the
initial hypersurface. The shadows are very clumped, gathering around
the rare regions where inflation continues longest, with an
arbitrarily large number of arbitrarily small overlaid shadows
surrounding the set (of measure zero) of points on the surface where
inflation continues forever. Thus, all counts are infinite
numbers and require regularization to be well-defined. The authors
propose only counting shadows larger than a size $\epsilon$ and then
taking the limit $\epsilon \rightarrow 0$. This measure is argued to
be independent of initial conditions on the surface and applies to
terminal and recycling vacua. It also has the important feature of
giving metastable states non-zero weight. While the idea of ``counting bubbles at future infinity" is intuitively clear, it is somewhat unclear that the ``shadow counting" used to actually implement the cutoff is particularly physical.

Moreover, converting this idea into an actual calculation is a subtle matter. To date, such calculations have been performed in a rate-equation framework in which one follows the
fractions of comoving volume in the various vacua and then effectively
``divides through'' by the bubble volume in order to obtain the bubble
count. The shadow-size cutoff is then implemented by imposing
a {\em set} of late-time cutoffs, one for each bubble type out of
which the counted bubbles are nucleated (on the assumption that this determines the ``comoving size" of the nucleated bubbles, and thus the size of the shadow, to which the cutoff applies). This cutoff, $t^{(\epsilon)}_{ij}$, for transitions out of vacuum $j$ into vacuum $i$, is given by~\cite{Garriga:2005av}
\begin{equation}
t^{(\epsilon)}_{ij} = -\ln\left(\epsilon H_{j} \right),
\end{equation} 
and is designed so that when bubbles intersecting the cutoff surface are projected back onto the initial surface, only bubbles of size exceeding $\epsilon$ will be obtained. 

There are, however, some features of this calculation that warrant a closer look. For example, the formalism allows situations in which a bubble formed soon after its parent can be assigned a larger asymptotic comoving size than the parent (we thank Alex Vilenkin for discussions of this point) and may therefore be included in the counting while its parent is not.  It is, however, unclear if or how the nucleation of bubbles larger than their parent actually occurs, or what asymptotic size should really be assigned to them. One might hope that such events lead to a small error, but this is not clear because the ratio of 4-volume between the cutoff surfaces to the full 4-volume before the cutoffs may be large. 
Thus rather than the time period between the cutoffs being unimportant, nucleations during this period may actually dominate the bubble statistics. Details of this sort should serve to encourage the development of calculational techniques in which spacetime dependence is more explicitly taken into account.

\item {\em The Worldline (W) method}: \ \ Easther
et al.~\cite{Easther:2005wi}, whose measure we denote the Worldline
(W) method, assume that at some initial time (defined by a spacelike
hypersurface), the universe is in some places in a non-terminal
vacuum. They then suggest considering a finite number of randomly
chosen points on this initial data surface and following forward
worldlines with randomly chosen velocities~\footnote{It is unclear the extent to which the velocities of individual points can be chosen at random, as discussed by Vilenkin in~\cite{Vilenkin:2006xv}} from these initial data
points. Only bubbles that are encountered by at least one
of these worldlines are counted in determining the relative bubble
abundance (no bubble is counted more than once, even if multiple
worldlines enter it).  One then takes the total number of worldlines
to infinity. Like CHC, this measure is claimed to be essentially
independent of initial conditions as long as inflation is eternal. It was argued in~\cite{Garriga:2005av} that the CHC and W methods of bubble counting
yield identical answers for terminal landscapes (the W method is ill-defined for fully recycling landscapes as discussed in~\cite{Vanchurin:2006qp}).

\end{itemize}

The remaining two measures focus on the transitions between vacua
experienced by a single eternal worldline, and accord a probability to
a vacuum that is proportional to the relative frequency with which it is
entered ($p$=``segment of a worldline between vacuum transitions").

\begin{itemize}

\item {\em The Recycling Transition (RT) method}: \ \ The proposal of
Vanchurin and Vilenkin~\cite{Vanchurin:2006qp}, which we will refer to
as the Recycling Transition (RT) method, is to follow the evolution of
a given geodesic observer and set the probability to be in a given
vacuum proportional to the frequency with which this vacuum is
entered, in the limit where the proper time elapsed goes to
infinity. As presented, the method only applies to landscapes with no terminal vacua, and was argued to be equivalent to the CHC
method in that case~\cite{Vanchurin:2006qp}.

\item {\em The Recycling and Terminal Transition (RTT) method}: \ \ The
Bousso proposal~\cite{Bousso:2006ev}, which we denote the Recycling
and Terminal Transition (RTT) method, covers the cases of terminal and
recycling vacua. Here, one chooses an initial condition for the
worldline (the predictions of this measure are dependent on initial
conditions), and considers the relative probabilities of the worldline
entering various other vacua, averaging over possible realizations.
This is equivalent to the RT measure in the case where there are no
terminal vacua.
 
The focus in RTT on the worldline of an observer is presented as being motivated by holography and the desire to only
consider regions of spacetime that an observer can signal to and
receive signals from (the ``causal diamond'').  However, this
viewpoint makes essentially no difference to the mathematics and -- as
mentioned below -- the time average over histories for Bousso's
observer could equally well be thought of as spatial averages over
widely-separated worldlines in any of the above approaches. A similar observation is made in~\cite{Linde:2006nw}. Of course, a holographic point of view might lead one to strongly disfavor further possible weighting factors to apply such as volume weighting.

\end{itemize}

Although we will not treat them further, let us also mention some
other approaches to asking about predictions in eternal inflation.
In~\cite{Tegmark:2004qd}, Tegmark advances a simple and direct
possible answer to the question of the relative numbers of different
vacuum regions: because eternal inflation should produce a countably
infinite number of each type of vacuum region, and because all
countable infinities are equal in the sense of being relatable by a
one-to-one mapping, each vacuum should be assigned equal weight.
In~\cite{Gibbons:2006pa}, the authors put a measure on the space of
classical FRW solutions to the Einstein plus scalar field
equations. If this could be extended to allow for quantum jumps
analogous to bubble nucleations, it might help address the
distribution of vacua within and amongst solutions.  In~\cite{Gratton:2005bi}, the authors focus on histories that might be/might have been observed, in the context of single-field inflation with a monotonic potential. 

\subsection{Relations between the measures}
\label{sec-relations}

Although the methods, both in their motivation and in their
presentation here, have been categorized into ``volume counting",
``bubble counting" and ``worldline following", there are relations
between them that cross these divisions, so that in fact there are
actually very few essentially different measures under consideration.

Some of the relations between measures (as presented by their authors)
have been mentioned above (e.g.\  the equality of CHC and W for terminal landscapes, and the equality of CHC and RT for ``fully recycling" landscapes with no
terminal vacua).  More, however, exist.

In particular, the RTT method accords the same relative probabilities
to terminal vacua as does the CV method (though the methods differ for
non-terminal vacua, which have zero probability in CV and nonzero
probability in RTT). To see this, consider a congruence of comoving
worldlines starting in some vacuum. Now, as $t\rightarrow\infty$,
every worldline that will eventually end up in a terminal vacuum will
do so (by definition); moreover, each terminal vacuum will only be
entered once (also by definition). Since RTT accords relative
probability to two terminal vacua A and B equal to the relative
probability of a worldline entering them, this will be equal
to the relative numbers of worldlines terminating in A versus B, which
is in turn equal to the relative $t\rightarrow\infty$ comoving volume
fractions as defined in the CV method. In appendix~\ref{snowmensection}, we show this correspondence by directly comparing the results of the RTT and CV methods in the context of a specific model. More generally, the results of the RTT method, for terminal as well as recycling landscapes, can be obtained by integrating the incoming probability current into the various vacua~\cite{Linde:2006nw,Garriga:1997ef}. 

\begin{figure}
\includegraphics[scale=0.45]{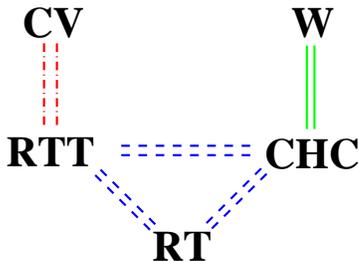}
\caption{
  \label{fig-connections}
A summary of the connections between the various measures. Solid green
lines indicate equivalence between the measures for a terminal landscape. Dashed blue lines indicate equivalence in the case of a fully
recycling landscape. Dashed-dotted red lines indicate that the measures assign
the same relative weights to terminal vacua.}
\end{figure}

These relations between the measures (as formulated in the original
papers) are summarized in Fig.~\ref{fig-connections}.  It also appears
possible to use what is understood about these connections to devise
some hybrid or generalized versions of the methods.

For example, take the CV procedure, where only a single late-time
hypersurface is considered, and attempt to count the number of bubbles
intersecting this surface from the volume distribution and some
appropriately defined cutoff. This is {\em not} quite the CHC method
since, as described above, the CHC calculation requires a different
time cutoff for bubbles formed in different parent vacua. But this
CV-CHC ``hybrid" prescription does not seem any less reasonable to
us. One could also generalize the CHC prescription to obtain an
infinite number of related measures by altering the limiting
procedure: rather than only counting shadows larger than a size
independent of the bubble type, one could instead only count shadows
larger than a given size relative to, say, some function of their
Hubble radius. That is, for bubbles of type $P$, rather than only
    counting those that have shadows larger than $\epsilon$ on the initial surface, count those that have shadows larger than $\epsilon' H_P$ or $\epsilon'' / H_P$ say.  This would correspond to replacing the time cutoff of $-\ln (\epsilon H_M)$ for bubbles of type $P$ forming out of bubbles of type $M$ with $-\ln (\epsilon' H_M H_P)$ or $-\ln(\epsilon'' H_M / H_P)$.
It  would be interesting to investigate how (in)sensitive the probabilities are
to the choice of a particular cutoff procedure. 

Having described the various bubble counting measures and their
connections, we now use a set of sample landscapes to illustrate some
of their predictions.

\section{Some Sample Landscapes}\label{samples}
\begin{figure}
\includegraphics[scale=0.5]{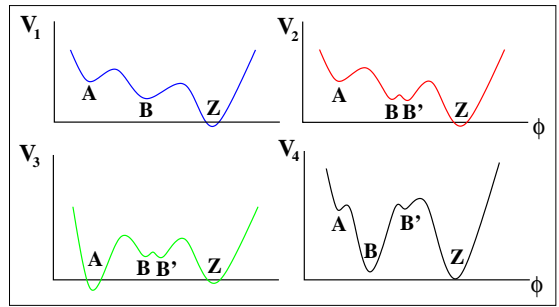}
\caption{
  \label{potentials}
Some sample landscapes. Potential $V_{1}$ depicts the $ABZ$ example
discussed by Bousso~\cite{Bousso:2006ev}. $V_{2}$ splits the B vacuum
by introducing a small barrier. Potential $V_{3}$ lowers the $A$
vacuum to zero or negative energy, so that it becomes terminal. The
potential $V_{4}$ has a low energy minimum with high-energy neighbors
that have short lifetimes (relative to other vacua in the
landscape).}
\end{figure}

Consider the related one-dimensional landscapes pictured in
Fig.~\ref{potentials}. They all contain both terminal and recycling vacua (where we assume here that a vacuum is terminal if and only if its energy is zero or negative), and we now discuss the predictions made by the RTT method for each. In light of the close connections between the measures, many of the conclusions drawn from these 
calculations will hold more generally.   

Following Bousso, we define the relative probability $\mu_{NM}$ to
transition from vacuum $M$ to vacuum $N$ as
\begin{equation}
\mu_{NM} \equiv \frac{\kappa_{NM}}{\sum_{P} \kappa_{PM}}
\end{equation}
where $P$ is summed over all decay channels out of $M$, and
$\kappa_{NM}$ is the probability per unit time of tunneling from
vacuum $M$ to vacuum $N$. Note that all summations in this paper are expressly indicated. $\kappa_{NM}$ typically takes the
form of a three-volume times a nucleation rate per unit four-volume,
the latter being calculated using semiclassical instanton techniques.
Note that
$\sum_P \mu_{PM}=1$ if $M$ is metastable and $\mu_{PM}=0$ if $M$ is
terminal, and also that $\mu_{MN} \neq \mu_{NM}$ in general.
Bousso introduces the concepts of trees and pruned trees in order to
calculate the prior distribution in the RTT method. He also presents a
matrix formulation, which we develop further in appendix~\ref{snowmensection}.

It will be important for what follows to obtain an indication of the
magnitudes of tunneling rates in a typical landscape. We model this
landscape by a single scalar field $\phi$ with a potential
$V(\phi)$ expressed as $V(\phi)=\mu^4 v(\phi/m)$. We further assume that
$v$ is a smooth function that varies over a range of order unity
as its argument changes by order unity, and $\mu$ sets the energy
scale. For the semi-classical approximation that we are working in to
make sense, we must have $\mu^4 \ll M_{\rm
  Pl}^4$, where $\mpl$ is the Planck Mass.  For Coleman--De Luccia
instantons to exist, $m$ must be less than some $O(1)$ multiple of $\mpl$.
See~\cite{Aguirre:2006ap} for more on the motivation for 
this form of the potential. 

As mentioned above, we will estimate tunneling
rates between the potential minima using semiclassical instanton
techniques, notwithstanding thorny issues of interpretation,
particularly for upward transitions.  
Then $\kappa_{NM} \propto e^{-(S_{(NM)}-S_{M})}$, the bracketed
exponential factor being the difference between the action $S_{(NM)}$
of the Coleman-De Luccia or Hawking-Moss
instanton linking the two vacua and the  
action $S_{M}$ of the Euclidean four-sphere corresponding to the
tunneled-from spacetime. Note that the same instanton applies to
uphill and downhill transitions (hence the use of symmetrising
brackets in its label). Using 
the Euclidean equations of motion, $S_{(NM)}$ can be written as
\begin{equation}
\label{sinst}
S_{(NM)} = -\int \sqrt{g} \, V(\phi) \, d^4 x
\end{equation}
where the integral is performed over the Euclidean manifold of the
instanton. The background subtraction term (which is negative and larger in magnitude than the instanton action) is given by the same expression and evaluates to
\begin{equation}\label{SBG}
S_{M} = - \frac{3 \mpl^4}{8 V(\phi_M)},
\end{equation}
where $ V(\phi_M)$ is the value of the
potential of the pre-tunneling vacuum $M$ at $\phi=\phi_M$.  

From these formulae we can immediately deduce two important facts.
First, we can compare uphill and downhill rates between two vacua.  In
the ratio of the rates the instanton part cancels out, and only the background parts are left. If $V(\phi_M) = V(\phi_N)+\Delta V$, then 
\begin{equation}
\frac{\kappa_{MN}}{\kappa_{NM}} \sim
\exp \frac{-3 \mpl^4}{8} \frac{ \Delta V }{ V^2(\phi_M)} = \exp \frac{-3}{8}
 \frac{\Delta v}{v_M^2} \left(\frac{\mpl}{\mu}\right)^4. 
\end{equation}
So, unless $\Delta v$ is tuned to be much smaller than $v$, the
uphill rate is exponentially smaller than the downhill rate. 

Second, we can compare the rates to two vacua $N$ and $P$ from the
same parent vacuum $M$.  This time the background parts cancel and we are
left with the exponential of the difference of the instanton actions:
\begin{equation}
\frac{\kappa_{PM}}{\kappa_{NM}} \sim \exp{-(S_{(PM)}-S_{(NM)})}.
\end{equation}  
Both instanton actions will be of order $(M_{\rm Pl}/\mu)^4$, 
 so we typically expect the tunneling rates to differ exponentially.  
In particular, if $V_N$ and $V_M$ are somewhat atypically similar and there is only a small barrier between the two, then, as long as $V_P$
is not atypically close to $V_M$ also, tunneling from $M$ to $P$ will be
exponentially disfavored relative to tunneling to $N$. This holds even if the tunneling from $M$ to $N$ is uphill and that from $M$ to $P$ is downhill.  This difference in tunneling rates can be extreme: for a typical inflationary energy scale of $\mu \sim 10^{16}$ GeV,
$\kappa_{PM}/\kappa_{NM} \sim e^{-10^{12}}$.

\subsection{Coupled pairs dominate in terminal landscapes}\label{subdividing}

We begin by considering the potential $V_{2}$ depicted in
Fig.~\ref{potentials}. We assume that the barrier
separating $B$ and $B'$ is very small, so that rapid transitions occur between the two wells. 
Thus we take $\kappa_{B'B} \gg \kappa_{AB}$ and $\kappa_{BB'} \gg \kappa_{ZB'}$.  Using the results of appendix~\ref{snowmensection}, in the limit we obtain:
\ba \label{subdiv}
\begin{pmatrix} P^{A,B,B'}_{A} \\ P^{A,B,B'}_{B} \\ P^{A,B,B'}_{B'} \\ P^{A,B,B'}_{Z} \end{pmatrix} \propto
\begin{pmatrix} \kappa_{BB'} \kappa_{AB} \\ \kappa_{BB'} \kappa_{B'B}
\\ \kappa_{BB'} \kappa_{B'B} \\ \kappa_{B'B}\kappa_{ZB'}
\end{pmatrix}
\ea
where $P^{M}_{N}$ is the ``prior" probability of Eq.~\ref{eq-psubx} (with
subscript $p$ dropped) to be in the vacuum $N$, given an initial state
in vacuum $M$. A multiple superscript indicates that the
same distribution applies to the listed initial states for the
transition rates under consideration.

There are a number of interesting points to note here. First
\begin{equation}\label{ratio1}
\frac{P^{A,B,B'}_{B}}{P^{A,B,B'}_{A}} = \frac{\kappa_{B'B}}{\kappa_{AB}} \gg 1
\end{equation}
\begin{equation}\label{ratio2}
\frac{P^{A,B,B'}_{B}}{P^{A,B,B'}_{Z}} =
\frac{\kappa_{BB'}}{\kappa_{ZB'}} \gg 1. 
\end{equation}
These ratios hold independent of initial conditions. Vacuum
$B'$ is similarly weighted relative to $A$ and $Z$. We
therefore see that (as might be expected in a measure that counts transitions)
metastable vacua participating in fast transitions with their
neighbors are weighted very heavily. Such regions certainly exist in a
landscape with sufficient complexity, and it is these regions that the
prior distribution in the RTT method will favor.
From our above estimates of typical transition rates in regimes with
energies somewhat below the Planck scale, factors of order $e^{10^{12}}$
should be commonplace.

Of course, {\em arbitrarily} fast transitions between $B$
and $B'$ (which give arbitrarily high weighting to both vacua) are
unrealistic. In reality, bubble collisions will become important, and
at high enough nucleation rates there will be percolation. In this
limit, there should then be a transition to a treatment in terms of
field-rolling and diffusion. In this regard, it would be desirable to
treat field diffusion as described by the stochastic formalism and
bubble nucleation (with collisions taken into account) in a unified
way (see~\cite{Garriga:1997ef} for work in this direction).

Although the CHC measure is inequivalent to the RTT measure in
landscapes with terminal vacua, it (and hence the W method)
nevertheless gives similar qualitative predictions. We can see this
by analyzing the ``FABI" model of~\cite{Garriga:2005av},
which, in the limit where $\kappa_{B'B} \gg \kappa_{AB}$ and $\kappa_{BB'} \gg \kappa_{ZB'}$, gives the same ratios as Eqs.~\ref{ratio1} and~\ref{ratio2}. Thus the CHC and W proposals weight fast-transitioning states exponentially more than others in exactly the same way the RTT method does.
The weighting can easily be large enough to dominate any volume factors, which appear in the full probability defined using the CHC method~\cite{Garriga:2005av}, unless the number of e-folds during the slow-roll period after a transition
is extreme.

We have seen that pairs of vacua undergoing fast transitions in both
directions are weighted very heavily, but what about transitions that
are fast in one direction only? For example, consider $V_{4}$ in
Fig.~\ref{potentials}, where there are quick transitions into $B$, but
transitions out of $B$ are strongly suppressed. Requiring only $\kappa_{BB'} \gg \kappa_{ZB'}$ in the probability tables from appendix~\ref{snowmensection} yields: \ba\label{4wellb}
\begin{pmatrix} P^{A,B,B'}_A \\ P^{A,B,B'}_B \\ P^{A,B,B'}_{B'} \\ P^{A,B,B'}_Z \end{pmatrix} \propto 
\begin{pmatrix} \kappa_{BB'} \kappa_{AB} \\ \kappa_{BB'} \left(\kappa_{AB} + \kappa_{B'B} \right) \\ \kappa_{BB'} \kappa_{B'B} \\  \kappa_{B'B}\kappa_{ZB'}
\end{pmatrix}.
\ea
It is apparent that vacuum $B$ will be the most probable vacuum in
this sample landscape. 
The relative weight of $A$ to $B'$ is very sensitive to the details
of the potential since, as shown above, there is an exponential
dependence on the difference in instanton actions (which itself tends to be
quite large). In the absence of extremely fine-tuned cancellation in
this difference (which would be required to make $\kappa_{AB} \sim
\kappa_{B'B}$), one of the two will be vastly more probable than the
other.  We have already considered the case where vacuum $B'$ is much
more likely than vacuum $A$ with landscape $V_2$ above.  So the other generic  
alternative is for vacua $A$ and $B$ to have probabilities very
close to one-half, vacuum $B'$ to be exponentially suppressed and
vacuum $Z$ to be even more suppressed.  
 
These two examples together make it clear that in order to obtain the
large weighting observed for potentials $V_{2}$ and $V_{3}$, there
must be {pairs} of vacua which undergo fast transitions in {\em both}
directions.  This allows for closed loops that produce large numbers of
bubbles of each of the vacua in the pair; in such cases the probabilities of both vacua scale with the product of the transition rates between them.

\subsection{Coupled pairs dominate in cyclic landscapes}

As one might expect, the extreme weighting of coupled pairs persists
if we raise the height of the $Z$ well of $V_2$ in Fig.~\ref{potentials} so that it is no longer terminal. From the calculations in appendix~\ref{snowmensection}, we find:
\begin{equation}
\frac{P^{A,B,B',Z}_{B}}{P^{A,B,B',Z}_{A}} \simeq
\frac{\kappa_{B'B}}{\kappa_{AB}} 
\end{equation}
\begin{equation}
\frac{P^{A,B,B',Z}_{B}}{P^{A,B,B',Z}_{Z}} \simeq
\frac{\kappa_{BB'}}{\kappa_{ZB'}} 
\end{equation}
with the same results for the ratios of $P_{B'}$ in place of $P_{B}$ to $P_{A}$ and $P_{Z}$.
This is of special interest because
for cyclic landscapes the predictions of the RTT
method agree with those of the CHC and RT methods (see
Fig.~\ref{fig-connections}).  Thus all of these measures will weight
rapidly transitioning vacua heavily. 

\subsection{Splitting vacua}\label{splitting}

A closely related ``test" to which we can put the RTT method to is to
consider the situation where potential $V_{2}$ is obtained from
potential $V_{1}$ (The ``ABZ" example of~\cite{Bousso:2006ev}) by
inserting a small potential barrier in the middle (B) well.  The ratio
of weights in the $A$ and $Z$ wells in potential $V_1$ is given by:
\begin{equation}
\frac{P^{A,B}_{A}}{P^{A,B}_{Z}} = \frac{\kappa_{AB}}{\kappa_{ZB}},
\end{equation}
which can be found from the result of~\cite{Bousso:2006ev} by
substituting $\epsilon = \kappa_{AB} / \left(\kappa_{AB} +\kappa_{ZB}
\right)$ and $1-\epsilon = \kappa_{ZB} / \left(\kappa_{AB}
+\kappa_{ZB} \right)$. Now let us insert the barrier in such a way
that the transition rates into and out of the $A$ and $Z$ wells remain
unaffected.  After the insertion, the relative weights of vacuum $A$
and $Z$ (in potential $V_{2}$) are then found from Eq.~\ref{subdiv} to be
\begin{equation}
\frac{P^{A,B,B'}_{A}}{P^{A,B,B'}_{Z}} = \frac{\kappa_{BB'}}{\kappa_{B'B}} \frac{\kappa_{AB}}{\kappa_{ZB'}}.
\end{equation}
Now we can consider two cases. First, if there is no symmetry as $B$
is interchanged with $B'$, then we see that inserting the barrier has
changed {\em both} the absolute probabilities (which are now strongly
weighted toward $B$ and $B'$), and also the {\em relative} weights of
the other vacua. Second, if the problem is symmetric under
interchange of $B$ and $B'$ (so that $\kappa_{BB'} = \kappa_{B'B}$ and
$\kappa_{ZB'} = \kappa_{AB}$), then the relative weights of $A$ and
$Z$ are unaffected; however, the absolute weights of both are still
altered drastically by this decomposition of $B$ into two identical
vacua with fast transitions between them. This is somewhat disturbing, and again points to the need for a smooth connection between ``vacuum transitions" and ``field evolution."

\subsection{Continuity of predictions}\label{continuity}

The next landscape we wish to consider is one of the simplest imaginable -- just a
double well potential. In this example, the predicted ratio of weights in
vacuum $A$ to that in $Z$ (in the case of full recycling) is identical
for the CHC, RT, and RTT methods, with $P_{A} / P_{Z} = 1$,
independent of the relative lifetimes of the states. The ratio of
weights predicted by the CV method is~\cite{Vanchurin:2006qp} $P_{A}/
P_{Z} = (H_{A} / H_{Z})^4 e^{S_{A}-S_{Z}}$, where $H_{A,Z}$ is the
Hubble constant and $S_{A,Z}$ the entropy of vacuum $A$ and $Z$
respectively. The difference is due to the fact
that the CHC, RT, and RTT methods count the frequency of
transitions while the CV method weights according to the time spent in
a given vacuum~\cite{Vanchurin:2006qp}.

Now consider shifting the entire potential down, such that the lower
well becomes a terminal vacuum. The predictions of the CHC, RT, and
RTT methods will remain identical until the lower well is exactly
terminal, at which point the CHC and RTT methods (the RT method
breaks down when the lower well becomes terminal) predict $P_{A} = 0$,
$P_{Z} = 1$~\footnote{It is worth noting that that this is completely
independent of the ratio of the lifetimes of the states, which might
be arbitrarily large~\cite{Aguirre:2006ap}.}.  Were this a correct
description of relevant probabilities, it would be very important in
making predictions to know if the energy of a minimum were zero or
different from zero by one part in $10^{10^{100}}$. The CV method will
predict this distribution as well, but will approach it in a {\em
continuous} manner ($S_{Z} \rightarrow \infty $, sending the ratio
$P_{A}/ P_{Z}$ to zero). The predictions of the CV method are for this
reason much more robust under small changes of the potential.

One possible way to avoid this discontinuity might be to reverse the
order of limits $t \rightarrow \infty$ and $\kappa^{-1}_{AZ}
\rightarrow \infty$. All of the measures discussed in this paper take
the $t \rightarrow \infty$ limit first, but one could perhaps define a
measure where the duration in time is held finite while
$\kappa^{-1}_{AZ} \rightarrow \infty$. Applying this to the two-well
example, as the lifetime of the lower well goes to infinity, the
expectation value of the number of transitions observed would smoothly
go to zero.  Alternatively, it may be the case that there are no truly
terminal vacua (with strictly zero probability of being tunneled
from)~\footnote{For example, if the ``L" process describe below in
Sec.~\ref{observers} occurs, it might mediate transitions away from
negative or zero-energy vacua. A heuristic argument in favor of
tunneling from negative ``big crunch" vacua was given
in~\cite{Banks:2005ru}.  Finally, we note that after tunneling to a
negative vacuum, the spacetime is an open FRW model with energy
density. Thus there may conceivably be tunneling before the ``crunch"
even if such tunneling is impossible from pure AdS or Minkowski
space.}. Finally, it may be that there is simply something conceptually
flawed in the way bubble-counting measures treat the borderline
between a vacuum being terminal and non-terminal.

\section{Consequences for predictions in a landscape}\label{consequences}

The previous section pointed out some interesting features of
bubble-counting measures (all the measures here save CV) as somewhat
abstract procedures applied to small ``toy" landscapes.  What might these
features imply for predictions (in the form of $P_p$ or ${\cal P}_X$) in
a more realistic landscape with many, many vacua and transitions connecting them?

Without a well-specified model of such a landscape this is a difficult
question to answer; however the strong preference for pairs of
fast-transitioning vacua does suggest some general -- and possibly
troubling -- predictions.  Within a landscape, imagine the set of all
pairs of neighboring vacua $(M,N)$ with similar pairs of energies
$(V_M,V_N)$, and suppose that for each pair, the barrier between $M$
and $N$ is independent of the barriers separating $M$ and $N$ from
other nearby vacua. Then we might expect that members of different pairs will be accorded exponentially differing 
probabilities depending on the details of the barrier.  
In Sec.~\ref{samples} we found in our sample
landscapes that the probabilities for the vacua in a fast-transitioning pair
$(N,M)$ are approximately proportional to the product
$\kappa_{NM}\kappa_{MN}$ of the transition rates between them.  What
determines this product?  We fix $V_M$ and $V_N$, and imagine
the possible potentials $v$ in-between (i.e.\ 
consider we consider many pairs in the landscape).  We have
\begin{equation}\label{kapproduct}
\kappa_{MN} \kappa_{NM} \sim e^{-2 S_{(MN)}(v)} e^{S_{M}+S_{N}},
\end{equation}
where $S_{(MN)}(v)$ is the instanton action of Eq.~\ref{sinst} and $S_{M,N}$
are the background subtractions for vacua $M$ and $N$, given by
Eq.~\ref{SBG}. With $S_{M}$ and $S_{N}$ fixed, the product then
depends just on  $S_{(MN)}$.  As argued above, this action will be of order $(M_{\rm Pl}/\mu)^4$, and vary by order unity as the parameters governing
the potential $v$ are varied. Thus the weightings of the members of
each pair do appear to be exponentially sensitive to the shape of the potential
in-between.   

Now imagine that our vacuum is one tunnel away from one of the vacua with
energy $V_N$.   All other things being equal, we should be likely
to come from any given one according to its weight.  The evolution
towards our vacuum depends on the shape of the potential, and because
$v$ is smooth this will not be independent of the shape of the
potential between the endpoints of the instanton.   If an observable $\alpha$ depends on the shape of the potential as our vacuum is approached, then this raises the possibility of it having an exponentially varying prior over an
observationally relevant range. A good example might be the number of
post-tunneling e-folds, which might possess a prior exponentially favoring a
particular number.

One might hope to compensate the prior probabilities $P_p$ favoring
cosmologies unlike ours using a conditionalization factor $n_{X,p}$
that disfavors them (e.g.\ conditionalizing on the existence of a
galaxy). In some cases, this seems plausible.  For example, if we
consider the cosmological constant $\Lambda$ and (unrealistically)
assume that all other cosmological parameters stay fixed to our
observed values, then $n_{X,p}(\Lambda)$ decreases as an exponential
in $\Lambda/\xi^4 Q^3$, where $Q\sim 10^{-5}$ is the fluctuation
amplitude and $\xi\sim 10^{-28}$ is the matter mass per photon in
Planck masses (e.g.,~\cite{Tegmark:2005dy}). Because this scale is so
much smaller than the scale over which the parameters of the potential
vary (i.e. $\xi^4 Q^3 \ll M$), the exponential variations of
$P_p(\Lambda)$ are likely to be nearly constant over a range of order
$\xi^4 Q^3$, so $n_{X,p}(\Lambda)$ would be effective in forcing
${\cal P}_X$ to give most weight to a region of parameter space near
to what we observe~\cite{Weinberg:1987dv,Schwartz-Perlov:2006hi}. But in
other cases this is far from clear; for example, the number of
inflationary e-folds is determined by the {\em high energy} structure
of the potential at and near tunneling, and the number of e-folds is
linked to the field value to which tunneling occurs, which is in turn
linked to the instanton solution and hence the tunneling rate. Thus
$n_{X,p}$ and $P_p$ might easily vary over the same scale in the
parameters governing the landscape potential, and the
conditionalization may be ineffective at forcing ${\cal P}_X$ to peak
in the observed range.

\section{Observers in Eternal Inflation}\label{observers}

Measures relying on properties experienced by a local ``observer"
(generally equated with a causal worldline) require that observers can
actually transition between the different vacua. It is not, however,
clear that this is always the case. In~\cite{Aguirre:2005nt}, two of
the authors found that in semi-classical Hamiltonian descriptions of
thin-wall tunneling, there are always two qualitatively different
types of transitions described by the same formalism.

One, called the ``R" tunneling geometry, is a generalization of
Coleman-De Luccia~\cite{Coleman:1980aw}/Lee-Weinberg~\cite{Lee:1987qc}
(CDL/LW) true and false vacuum bubbles. It corresponds to the
fluctuation of a bubble of the new phase which is always in causal
contact with the background region, in the sense that worldlines in
the old phase can both ``tunnel with" the bubble, and also enter the
bubble of new phase soon after it forms.

In the other, which was called the ``L" tunneling geometry (a
generalization of the Farhi-Guth-Guven mechanism~\cite{Farhi:1989yr}),
the bubble of new phase lies behind a wormhole separating it from the
original background spacetime. In this case, no causal curve from the
original phase can enter the new phase after the tunneling event (in
marked contrast to the usual picture of an expanding bubble of new
phase, or to the R mechanism). Some rare worldlines might ``tunnel
with" the bubble, but the physical connection between pre-and
post-tunneling phases represented by such worldlines is obscure at
best; moreover such worldlines do not exist in the (highest probability)
limit in which the bubble has zero mass.

If both L and R processes occur, then the L mechanism is the most
probable path by which regions of {\em higher} vacuum energy
emerge, while the R geometry dominates decay to a lower
vacuum~\cite{Aguirre:2005nt}; both processes are dominated by the
lowest-mass bubbles.

At the semi-classical level of these calculations, the authors
of~\cite{Aguirre:2005nt} found no convincing reason that one but not
the other of these two tunneling processes would occur. Holographic
considerations would seem to conflict with the L geometries (at least
for transitions to higher vacuum energy), and~\cite{Freivogel:2005qh}
argued using AdS/CFT that such events tunneling from AdS to dS would
correspond to non-unitary processes; however the question has not been
settled with any clarity. (See~\cite{Alberghi:1999kd} for another treatment of L tunneling geometries using AdS/CFT.) In this section we will therefore consider
how the L-tunneling process would impact eternal inflation, and the
measures as applied to it.

\begin{figure*}
\includegraphics[scale=.35]{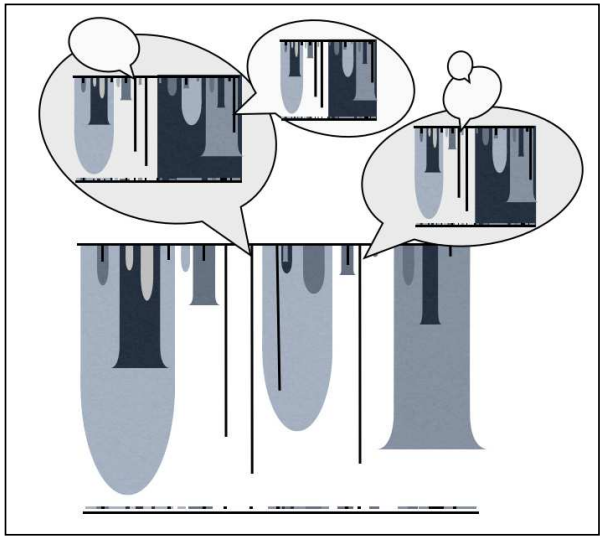}
\caption{
  \label{landruniverse}
A picture of an eternally inflating universe which takes into account
both L and R tunneling geometries. At the bottom, there is an
``original'' parcel of comoving volume (defined by the horizontal
spacelike slice at the bottom of the figure), which evolves in time
(vertically). True and false vacuum bubble nucleation events occur via
the R geometry in this volume, denoted by the shaded regions which in
the case of true vacuum bubbles grow to a comoving Hubble volume and
in the case of false vacuum bubbles shrink to a comoving Hubble
volume. The vertical black lines denote the black holes formed during
L geometry tunneling events. On the other side of a wormhole (inside
the captions), the initial distribution, which is fixed by the
tunneling geometry, undergoes L and R tunneling events as well,
spawning more disconnected parcels of volume in which this process
repeats. The original parcel of comoving volume will spawn an infinite
amount of new comoving volume via L geometry tunneling events. Shown
on the bottom of each parcel is the set of bubble shadows that might
be used in the CHC method to calculate probabilities $P^{{\cal V}_i}$
for each region ${\cal V}_i$.}
\end{figure*}

Let us consider an initial parcel of comoving volume in a metastable state residing in an arbitrary potential landscape. This is shown at the bottom of Fig.~\ref{landruniverse}. As time goes on, bubbles of either higher or lower vacuum energy will nucleate by either the L or R tunneling geometries. Since low-mass bubbles are most probable, most downward transitions will be CDL bubbles (the R geometry in the zero mass limit), and most upward transitions will be L-geometry tunneling events corresponding to a very small mass black hole forming in the background spacetime. Such small black holes affect the background spacetime in a completely negligible way as long as the nucleation rate is rather small~\footnote{In fact even more probable is the zero-mass limit in which there is no black hole at all, which also clearly does not affect the background spacetime.}.  In particular, these upward nucleations remove zero comoving volume from the old phase.

The pre-and post-tunneling spacetimes in an L-tunneling event are
described comprehensively in, e.g.,~\cite{Aguirre:2005nt}; the portion of the 
post-tunneling spacetime existing behind the wormhole consists of regions with both new and old vacuum energy separated by a thin wall, and in the zero-mass limit is just the Lorentzian CDL bounce geometry. Both vacuum regions are
larger than their corresponding Hubble radii and so will unavoidably
continue to inflate, independent of the precise details of the initial
nucleated space (i.e.\ how the instanton is ``sliced" to be continued
into Lorentzian space; see~\cite{Garriga:1997ef} for the corresponding
issue concerning the CDL instanton).
 
The result is that an entirely new ``branch" of eternal inflation is
  created, with some initial physical volume, having essentially no
  effect on the original spacetime. If a comoving volume is assigned
  to this physical volume using the ``scale factor time" of the
  background geometry near the nucleation event, then the effect will
  be to create {\em new comoving volume}~\footnote{How to actually
  define ``comoving volume" in the new phase is very unclear; comoving
  volume is related to a particular coordinatization of a spacetime,
  and its definition is tied to a congruence of geodesics; here no
  such congruence continues through the nucleation to fill the initial
  slice.}.  The new branch will in turn spawn more branches -- and
  more comoving volume -- via L-events, so that the comoving volume
  appears to actually grow exponentially (though in what ``time" this
  occurs is unclear since there is no foliation of the entire
  spacetime). This process is shown in Fig.~\ref{landruniverse}.

How do the measures we have been discussing connect with this new
picture? Consider first the measures RTT, RT and W that explicitly
follow causal worldlines.  As formulated, these measures would
essentially ``ignore" L transitions.  This seems quite artificial,
however, as regions with high vacuum energy (reached by upward
transitions) would almost all arise from this process; put another
way, choosing a random point in the entire spacetime (including the
tree of new universes formed by the L tunneling geometry) and
projecting any geodesic back, it would almost certainly hit an
L-geometry nucleation surface in the past rather than the assumed
initial slice.

Now consider the CV and CHC prescriptions. As stated, the idea is to
count the relative comoving volume or number of bubbles of different
types ``on future null infinity".  But as described in
Sec.~\ref{measprops} and in~\cite{Garriga:2001ri,Garriga:2005av,Vilenkin:2006xv}, these measures are actually {\em calculated} with very strong reliance on a
congruence of geodesics emanating from an initial surface; thus as calculated in this formulation they would be as unaffected by L-geometry events as
RTT, RT, and W. It is interesting, however, to speculate about taking
these prescriptions seriously as counting bubbles on future infinity,
as this would actually include the bubbles in the other branches
created by L-events.

Consider, then, a volume ${\cal V}_{i}$ nucleated by an L-event (with
the subscript $i$ labeling the particular region under consideration),
and imagine a congruence of geodesics emanating from it, denoting by
${\cal J}^+({\cal V}_{i})$ the part of the spacetime's future null
infinity reachable by these geodesics. Then we might ``count bubbles
of comoving size exceeding $\epsilon$" (for CHC) or ``count comoving
volume" (for CV) on ${\cal J}^+({\cal V}_{i})$, to define a set of
relative probabilities $P^{{\cal V}_{i}}$.

Now, it is very unclear how precisely to combine the $P^{{\cal
V}_{i}}$ in all of the branches $i$ formed from L-tunnelings out of
both the original spacetime, and out of the future of ${\cal V}_{i}$,
and from the descendants of these branches, etc.  Nonetheless, some
general statements might be made even in the absence of such
precision.

Consider first CHC.  Since its probabilities are essentially
independent of ${\cal V}_{i}$, it seems that $P^{{\cal V}_i}$ will be
the same in all branches, so it is hard to see how anything else could
result from combining them.

Now consider CV, which {\em is} dependent on the initial conditions for ${\cal
V}_{i}$. Here, the ``initial" conditions for a branch are not provided
by the original spacetime, but rather by the dynamics of the
L-tunneling process, with a different set corresponding to each pair
of vacua between which the nucleations can occur. Whatever way we
calculate all of the $P^{{\cal V}_{i}}$, it seems likely that the
original spacetime's initial conditions will be completely overwhelmed
by those of all of the branches in the infinite self-similar tree
depicted in Fig.~\ref{landruniverse}. One might then imagine that the
total prior distribution $P$ is given by a weighted sum of these
separate distributions, and is independent of the initial conditions
of the original spacetime.

We also point out that these questions may apply to ``stochastic"
eternal inflation as well. It is generally implicitly assumed in these
models that the global spacetime is causally connected, but this is
far from proven. Indeed, large fluctuations generically cause a large
back reaction, and it is not obvious that the large stochastic
fluctuations driving eternal inflation do not cause the production of
universes behind a wormhole (this is suggested by singularity
theorems~\cite{Farhi:1986ty,Vachaspati:1998dy,Aguirre:2005xs}). 
This discussion is also relevant for hypothetical transitions out of
negative energy minima. While no instanton has been constructed for
such a transition (see~\cite{Banks:2005ru} for a proposal concerning the probability of such a process), 
if one exists then (considering thin-wall
constructions~\cite{Freivogel:2005qh}) it would have to be an L
geometry. Based on the considerations above, it is unclear how or if
including such transitions would change the predictions of extant
measures.

\section{Discussion and Conclusions}\label{conclusions}

\begin{table*}[htbp] 
\begin{tabular}{|c|c|c|c|c|c|r|}
    \hline
Property & CV & CHC & W & RT & RTT  \\
    \hline
Physicality & P & P & P & P & P \\
Gauge independence   & P & Y & Y & Y & Y \\ 
Independence of initial conditions   & N & Y & Y  & Y & N \\
Copes with varieties of transitions and vacua  & P & Y & N & N & Y\\
Copes with nontrivial topologies  & P & P & N & N & N  \\
Treatment of states and transitions: &  &  & &  &  \\ 
-- General principles & P & P & P & P & P  \\
-- Physical description of transitions & P & P & P  & N & N \\ 
-- Reasonable treatment of split states & Y & N & N & N & N\\
-- Continuity in transition rates   & Y & N & N & N & N\\ 

    \hline
\end{tabular}
\caption{Properties of bubble counting measures -- Y=yes,
  N=no, P=partial.  \label{measures}} 
\end{table*}

We have analyzed a number of existing measures for eternal inflation,
exploring connections that exist between them, and highlighting some
generic predictions that they make. With this perspective, let us
return to the list of desiderata presented in
Sec.~\ref{wishlist}. Shown in Table~\ref{measures} is a ``scorecard"
detailing which of the measures, in at least a majority of the
authors' humble and irresolute opinions, satisfy the properties listed
in Sec.~\ref{wishlist}.

First, which measures are ``physical", in the sense of providing a
non-arbitrary prior probability $P_p$, for some ``counting object"
$p$, useful for calculating ${\cal P}_X$?  Physical volume weighting
(discussed little here) would seem quite physical but appears to lead
to gauge dependence~\cite{Winitzki:2005ya,Linde:1993xx}, and incorrect predictions in at least some gauges (see~\cite{Guth:2000ka,Tegmark:2004qd}).
The related CV ($p$=``unit of comoving volume") method may avoid some
of this difficulty, but at some cost to physicality: comoving volumes
are generally meaningful only insofar as they are re-converted to
physical ones, or if there are conserved objects (baryons, galaxies,
etc.) with fixed density per unit comoving volume. The latter may be
true after reheating, but it is unclear to us that comoving volume is
as meaningful {\em during} a complex, inhomogenous inflationary
period. Another option is to weight according to the integrated incoming probability current~\cite{Garcia-Bellido:1993wn,Linde:2006nw} across reheating surfaces, which can be found directly from volume distributions. This proposal, which is tied more closely to the conditionalization, avoids the gauge dependence and spurious predictions of standard volume weighting (as discussed above, this prescription can reproduce the results of the RTT method~\cite{Linde:2006nw,Garriga:1997ef}). 

The CHC and W methods have $p=$``bubbles,'' which might be
tied to conditionalization objects associated with the various
reheating surfaces (though this involves considerable uncertainty
since those reheating surfaces are generically infinite). However, the
objects (worldlines and shadows) actually used to arrive at a bubble
count seem rather less physical, particularly as they demand a cutoff
prescription that -- while natural -- also seems as if it could easily
be different.
The RT and RTT methods use $p=$``segment of a worldline between vacuum
 transitions,'' and has been suggested as an appropriate measure if we
 identify $X=$``unit of entropy
 production''~\cite{Bousso:2006ev,Bousso:2006xc}. This connection is
 not entirely compelling, however, as the results of these
 ``holographic'' measures can be found by considering an ensemble of
 observers (as noted in Sec.~\ref{measprops} and
 by~\cite{Linde:2006nw}). These connections suggest that CV, RT, and
 RTT are very closely related, but with a consistent and appropriate
 physical interpretation somewhat lacking.

Consider now gauge independence. Physical volume weighting is gauge
dependent, but the other measures appear gauge-independent, albeit
with some caveats. For RT, RTT, W, and CHC, gauge-independence stems from
their counting of objects (bubbles) or events (transitions); in CV it
occurs via use of a congruence of geodesics, which are also then
``counted" to obtain comoving volume.  The caveats stem from
subtleties -- connected with a time variable choice -- in defining
cutoffs, transitions rates, and initial conditions, and we hope to
elucidate some of these further in future work. (We single out CV as
partially gauge-dependent because the results will depend on the time
slicing used to characterize the initial value surface.)

Drawing on the description of the various measures presented in Sec.~\ref{measprops}, we can see that not all of the measures under discussion have the ability to cope with all types of transitions and vacua. For instance, the CV method accords zero weight to metastable minima (particularly disturbing as we may live in one), and the RT method in its current formulation is not able to describe a landscape with terminal vacua. We also note that the CV and RTT methods are dependent on initial conditions. 

In Sec.~\ref{observers}, we argued that it is possible -- if certain
types of ``L" bubble nucleation events occur -- for different regions
of the eternally inflating multiverse to be separated by wormholes,
and therefore causally disconnected. None of the evaluated measures
are, as formulated, equipped to deal with such spacetimes in a
reasonable way.  The ``philosophy" behind CV and CHC -- of counting
bubbles or volume on future infinity -- might reasonably apply to such
spacetimes, and if this could be implemented technically we argued
that in this case CV would probably become independent of initial
conditions. The philosophy behind RTT and RT would suggest simply
ignoring these events (as indeed those measures effectively do) but it
is rather unclear to us that this is appropriate.  Accounting for such
tunneling events in measure prescriptions is very difficult -- but
this merely highlights the possible importance of such transitions, and of
determining whether or not they occur.

Even thornier problems might arise from considering transitions in
 greater generality.  All of the measures considered rely on a
congruence of worldlines and a fairly straightforward spacetime
structure. Were we to include transitions between different string/M
theory flux vacua, including even different numbers of large spacetime
dimensions, it is unclear whether the principles of extant measures
would apply. Without having a well--defined description of such
transitions this is difficult to asses, hence we do not consider this
in our table.

But even confining our attention to (relatively) well-understood
spacetime evolution in a general scalar potential landscape, the
measures differ somewhat in how generally and robustly they treat
``vacua" and ``transitions".  All of the measures under discussion
have been applied to the brand of eternal inflation driven by
metastable minima. However, it would be desirable to include the
effects of all the dynamics of an eternally inflating universe, and
the effective scalar fields that are imagined to drive it. This
includes a description of the diffusion and classical rolling of the
field that will occur. There has been work extending CV and CHC
methods to these cases, but little so far in making such an extension
to RT or RTT.

In terms of connecting transition rates to physical transitions, all
of the measures ignore the small-scale details of vacuum transitions
(i.e. within a few Hubble volumes).  This may be relatively benign,
but bears investigation.  For example in RTT ``transitions" are
thought of as something that occurs to a worldline within its causal
diamond -- but these transitions could occur via the encounter of a
bubble formed in a nucleation process {\em outside} the causal
diamond.

More trouble occurs when we consider nearby vacua separated by a small
barrier.  The main observations of this paper centered around a study
of the sample landscapes shown in Fig.~\ref{potentials} using the RTT
method. In Sec.~\ref{subdividing} it was found that pairs of vacua
that undergo fast transitions will be very strongly weighted. Using
order of magnitude estimates of the transition rates, we argued that
the probability ratio of such pairs to other vacua in the sample
landscape can be exponentially large. This effect occurs in both
terminal and recycling landscapes. Using the equivalences between the
various measures noted in Sec.~\ref{sec-relations} (for a summary, see
Fig.~\ref{fig-connections}), and an explicit example for the CHC
method, we have shown that the weighting of fast-transitioning pairs
occurs in the CHC, W, and RT methods as well.  As discussed in
Sec.~\ref{splitting}, because of this effect, by inserting a
small barrier in an intermediate state, the absolute weight assigned
to each vacuum is affected drastically. Therefore, the RTT, RT, W, and
CHC methods are only partially robust in their definition of
transitions; the undivided-well distribution is not recovered as the
barrier disappears. This situation might be remedied if, as bubble
collisions become more and more important, the diffusion analysis
replaces bubble nucleation (giving further impetus to generalizing the
measures to treat this). In contrast, the CV method {\em does}
approach the undivided-well weight as the small barrier disappears.

Lastly, we considered continuity in transition rates, which was studied
using a two-well landscape in Sec.~\ref{continuity}. It was noted that
the predictions of the CHC, RT, and RTT methods change
discontinuously as a recycling vacuum is deformed into a terminal
vacuum. This discontinuity makes the {\em exact} properties of vacua
in a landscape important. Such a discontinuity could potentially be avoided if the
order of limits in the cutoff procedure were modified.

Most of the discussion -- and all of the scorecard -- has focused on
issues of principle concerning the measures as abstract procedures.
Some of the discussed features have implications for what such assumed
measured would mean {\em observationally}.  In particular, we saw in
Sec.~\ref{consequences} that the exponential dependence of the prior
distribution $P_p$ on the details of the potential implies that making
predictions using bubble counting measures may be very hard. This
problem is particularly acute when, for some parameter $\alpha$, the
factors $P_{p}(\alpha)$ and $n_{X,p}(\alpha)$ (these are the prior and
conditionalization factors needed to produce a prediction in the form
of Eq.~\ref{eq-psubx}) vary appreciably over the same range in
$\alpha$.  This may be the case, for example, when $\alpha$ is related
to the number of e-folds during inflation.  If the observation that
fast-transitioning pairs are exponentially weighted generalizes to
more complicated landscapes, then bubble-counting measures may in some
cases lead to strongly exponential prior probabilities that would
overwhelm any conditionalization factor $n_{X,p}(\alpha)$.  This would
lead to very strong predictions, which might be successful, or
disastrous.  More generally, this exponential dependence suggests that current measures seem to potentially call for a complete knowledge of the fine details of the entire landscape, a Herculean requirement.

Perhaps not surprisingly, we come to the conclusion that while
progress has been made towards predicting our place in the multiverse,
we are far from finished. 
It would be desirable to find and explore
other measures, and see if they fall victim to any of the same
problems that we have outlined.

\begin{acknowledgments}

We thank Raphael Bousso, Ben Freivogel, Antony Lewis, Andrei Linde, Alex Vilenkin, and Sergei Winitzki for helpful discussions.  AA and MJ were partially supported by a ``Foundational Questions in Physics and Cosmology" grant from the
John Templeton Foundation during the preparation of this work. SG is
supported by PPARC. 
\end{acknowledgments}

\appendix

\section{Matrix Calculations and Snowman Diagrams}
\label{snowmensection}

In this appendix we present a quick way of calculating normalized
probabilities for terminal and cyclic landscapes in a unified manner,
which also sheds light on the nature of the regularizing limit taken
in the cyclic case.

First, assemble the relative transition probabilities
$\mu_{NM}$ into a matrix $\bm{\mu}$ (equivalent to Bousso's $\eta$
matrix).  Starting in an initial state represented by a vector $\bm{q}$
with components $q_N$ ($\Sigma_N q_N =1$), after one transition the mean
number of entries (or ``raw probability'') for each vacuum will be given
by $\bm{\mu} \bm{q}$.  At the second transition an additional
$\bm{\mu}^2 \bm{q}$ entries will occur and so on.  After $n$ transitions
the raw probability will be given by $(\bm{\mu}+\bm{\mu}^2 + \ldots
\bm{\mu}^n ) \bm{q}$.  If we set $\bm{S}_n \equiv \bm{\mu}+\bm{\mu}^2 + \ldots
\bm{\mu}^n $, then $(\bm{1}-\bm{\mu})\bm{S}_n=\bm{\mu}(\bm{1}-\bm{\mu}^n)$.  
In the terminal case we can invert $(\bm{1}-\bm{\mu})$ and take the $n\rar\infty$ limit  to obtain
$\bm{S}_\infty$ directly ($\bm{\mu}^n\rar 0 $ since asymptotically all the probability goes into the terminal vacua and so fewer and fewer vacuum entries occur).  In the cyclic case $\det(\bm{1}-\bm{\mu})=0$ and $\bm{\mu}^n$ does not tend to zero, and things are not so simple.  It is convenient to proceed by
replacing $\bm{\mu}$ by $(1-\varepsilon)\bm{\mu}$ (where $\varepsilon$ is an auxiliary parameter to be taken to zero after the calculation), which can be inverted. Neglecting the troublesome determinant factor (since we shall be later
normalizing to obtain probabilities from numbers of vacuum entries anyway), we take the limits $n\rar\infty$ and $\varepsilon \rar 0$ in that order, and for both terminal and recycling landscapes
 obtain the simple expression:
\ba
\bm{S}_\infty \propto \bm{T} \equiv \left( \text{adj} (\bm{1}-\bm{\mu}) \right) \bm{\mu}
\ea
where $\text{adj}$ denotes the adjoint matrix operation (i.e.\ the transpose of the matrix of cofactors of the matrix in question).  Multiplying
$\bm{T}$ into $\bm{q}$ and normalizing yields the probabilities for the vacua given the initial state in question.       

This procedure yields exactly the same results as the pruned tree
method.  We thus see that the latter procedure is equivalent to
considering sequences of transitions up to some length $n$ and then
taking the limit $n\rar \infty$.

\begin{figure}
\includegraphics[scale=0.28]{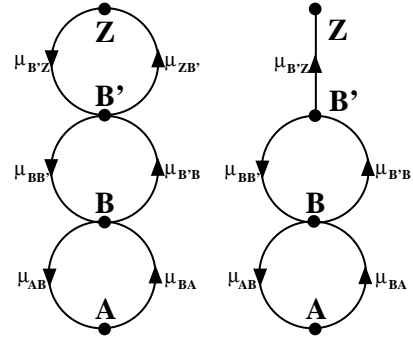}
\caption{
  \label{snowman}
Examples of ``snowman diagrams'' summarizing relative transition
probabilities $\mu_{NM}$.  The one on the left is for a recycling
landscape and 
the one on the right is for a terminal landscape.} 
\end{figure}

The $\mu_{NM}$s in question can conveniently be depicted in snowman-like diagrams such as those shown in Fig.~\ref{snowman}, which apply to the calculations in Sec.~\ref{samples}. These diagrams emphasize that the path between any two vacua can involve an arbitrary number of circulations in closed loops between recycling vacua. In fact we treat both cases at once by leaving $\mu_{ZB'}$ arbitrary and only set it to 1 or 0 as appropriate after having calculated $\bm{T}$.  We also allow for the possibility of vacuum $A$ being terminal in the same manner.

Suppressing the normalizing factor for clarity, we obtain 
\ba
\begin{pmatrix} P^{A,B,B',Z}_{A} \\ P^{A,B,B',Z}_{B} \\
  P^{A,B,B',Z}_{B'} \\ P^{A,B,B',Z}_{Z} \end{pmatrix} \propto
\begin{pmatrix} \mu_{AB} (1-\mu_{B'Z}\mu_{ZB'}) \\ 1-\mu_{B'Z}\mu_{ZB'}\\ \mu_{B'B} \\ \mu_{B'B}\mu_{ZB'} 
\end{pmatrix}
\ea
in the recycling case 
with the full set of superscripts indicating that the results are
independent of initial conditions. 

In the terminal case we can only start in states $A$, $B$ or $B'$ and
we obtain:
\ba\label{PA}
\begin{pmatrix} P^{A}_{A} \\ P^{A}_{B} \\ P^{A}_{B'} \\ P^{A}_{Z}
\end{pmatrix} \propto
\begin{pmatrix} \mu_{AB}  \\ 1\\ \mu_{B'B} \\ \mu_{ZB'}\mu_{B'B} 
\end{pmatrix},
\ea
\ba\label{PB}
\begin{pmatrix} P^{B}_{A} \\ P^{B}_{B} \\ P^{B}_{B'} \\ P^{B}_{Z}
\end{pmatrix} \propto
\begin{pmatrix} \mu_{AB}  \\ \mu_{AB}\mu_{BA}+\mu_{BB'}\mu_{B'B} \\ \mu_{B'B} \\ \mu_{B'B} \mu_{ZB'} 
\end{pmatrix}
\ea
and
\ba\label{PBP}
\begin{pmatrix} P^{B'}_{A} \\ P^{B'}_{B} \\ P^{B'}_{B'} \\ P^{B'}_{Z}
\end{pmatrix} \propto
\begin{pmatrix} \mu_{AB} \mu_{BB'} \\ \mu_{BB'} \\ \mu_{BB'} \mu_{B'B}
  \\ \mu_{ZB'} (1-\mu_{AB} \mu_{BA})
\end{pmatrix}.
\ea

The relative transition probabilities are related to the transition rates by
\ba
\mu_{BA} &=& 0 \text{~or~} 1 \\
\mu_{AB} &=& \frac{\kappa_{AB}}{\kappa_{AB} + \kappa_{B'B}} \\
\mu_{B'B} &=&  \frac{\kappa_{B'B}}{\kappa_{AB} + \kappa_{B'B}} \\
\mu_{BB'} &=&  \frac{\kappa_{BB'}}{\kappa_{ZB'} + \kappa_{BB'}} \\
\mu_{ZB'} &=&  \frac{\kappa_{ZB'}}{\kappa_{ZB'} + \kappa_{BB'}}
\ea
where $\mu_{BA} = 0$ if $A$ is terminal and $\mu_{BA} = 1$ if it
isn't. Substituting these expressions into
equations~\ref{PA},~\ref{PB}, and~\ref{PBP}, we can then take the
limits discussed in 
Sec.~\ref{samples} to produce the appropriate probability tables.  

In the case where vacuum $A$ is terminal ($\mu_{AB}=0$), there are a
number of ratios of interest. The probabilities assigned by the CV
method to this sample landscape were calculated
in~\cite{Garriga:2005av} (the ``FABI" model), and using these results,
we can directly compare the results of the CV and RTT methods. For
initial conditions in $B$ or $B'$, we find: 
\begin{equation}
\frac{P^{B}_{A}}{P^{B}_{Z}} = \frac{\kappa_{AB} \left(\kappa_{BB'} +
  \kappa_{ZB'}\right)}{\kappa_{B'B} \kappa_{ZB'}} 
\end{equation}
\begin{equation}
\frac{P^{B'}_{A}}{P^{B'}_{Z}} = \frac{\kappa_{AB}
  \kappa_{BB'}}{\kappa_{ZB'} \left( \kappa_{AB} + \kappa_{B'B}
  \right)}  
\end{equation}
As expected given the argument of Sec.~\ref{sec-relations}, these
results agree with the predictions of the CV method.

\bibliography{probet.bib}

\end{document}